# Electron density control in WSe$_2$ monolayers via photochlorination

E. Katsipoulaki[1,2,†], G. Vailakis[1,3,†], I. Demeridou[1], D. Karfaridis[4], P. Patsalas[4], K. Watanabe[5], T. Taniguchi[6], I. Paradisanos[1], G. Kopidakis[1,3], G. Kioseoglou[1,3,*], and E. Stratakis[1,2*]

[1]Institute of Electronic Structure and Laser, Foundation for Research and Technology-Hellas, Heraklion, 70013, Greece
[2]Department of Physics, University of Crete, Heraklion, 70013, Greece
[3]Department of Materials Science and Technology, University of Crete, Heraklion, 70013, Greece
[4]Department of Physics, Aristotle University of Thessaloniki, Thessaloniki, 54124, Greece
[5] Research Center for Electronic and Optical Materials, National Institute for Materials Science, 1-1 Namiki, Tsukuba 305-0044, Japan
[6] Research Center for Materials Nanoarchitectonics, National Institute for Materials Science, 1-1 Namiki, Tsukuba 305-0044, Japan

*E-mail: stratak@iesl.forth.gr and gnk@materials.uoc.gr
†These authors contributed equally

Modulation of the Fermi level using an ultraviolet (UV)-assisted photochemical method is demonstrated in tungsten diselenide monolayers. Systematic shifts and relative intensities between charged and neutral exciton species indicate a progressive and controllable decrease of the electron density and switch tungsten diselenide from n-type to a p-type semiconductor. The presence of chlorine in the 2D crystal shifts the Fermi level closer to the valence band while the effect can be only partially reversible via continuous wave laser rastering process. The presence of chlorine species in the lattice is validated by X-ray photoelectron spectroscopy (XPS), and density functional theory (DFT) calculations predict that adsorption of chlorine on the selenium vacancy sites leads to p-type doping. The results of our study indicate that photochemical techniques have the potential to enhance the performance of various 2D materials, making them suitable for potential applications in optoelectronics.

## I. INTRODUCTION

Monolayer (1L) transition metal dichalcogenides (TMDs) are materials with the formula MX$_2$, where M is a transition metal and X is a chalcogen atom [1]. Their electrical [1], optical [2,3] and mechanical properties [4] make them strong candidates for a variety of applications, such as field effect transistors, light-emitting diodes, photodetectors and photovoltaic solar cells [5–7]. Crucial for the performance of TMDs in relevant applications is the residual carrier density (or doping level), mainly originating from imperfections of the crystal, e.g. chalcogen or transition metal vacancies [8,9].

The density of charge carriers can affect a plethora of physical properties in TMDs, including their optical or electrical response. As a result, it is necessary to develop methods that can effectively control the doping level. This can be achieved either chemically or electrostatically. The former includes substitutional [10,11], charge transfer [12,13] or intercalation processes [14,15] while the latter can be realized by electrostatic field effect devices [16,17]. Another approach involves chemical processes assisted by light (i.e., photochemical) [18,19]. In particular, irradiation of TMDs with ultraviolet laser pulses in the presence of a gaseous precursor has the potential to effectively tune the carrier density of TMD semiconductors.

In this work, we apply a photochemical method to tune the optical response of mechanically exfoliated 1L-WSe$_2$. We irradiate our samples with 248 nm nanosecond pulses in the presence of chlorine gas. We perform photoluminescence (PL), Raman and differential reflectivity spectroscopy to monitor the impact of the variations in the carrier density after each photochlorination step. In low-temperature PL spectroscopy we observe a switch from negatively ($X^-_T$) to positively ($X^+_T$) charged trions, and a subsequent further enhancement in their intensity over neutral ($X^0$) excitons with additional photochlorination steps. XPS experiments confirm the presence of chlorine species in the monolayer lattice while DFT calculations predict p-type doping when chlorine binds with W in selenium vacancy sites. The results provide strong evidence that the photochlorination process is an effective means of controlling the carrier density and converting 1L WSe$_2$ from n-type to p-type semiconductor.

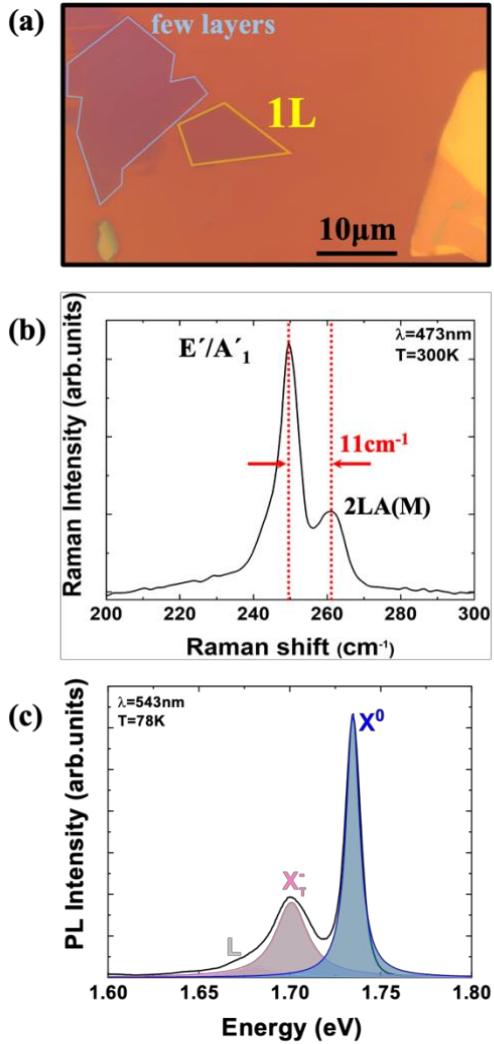

FIG. 1. (a) Optical image of 1L (yellow) and few layer (light blue) WSe$_2$ on SiO$_2$. (b) Typical Raman spectrum collected at T = 300 K with a 473 nm excitation laser line. (c) PL spectrum of 1L-WSe$_2$ at 78 K excited by a 543 nm continuous wave laser. Emission from neutral (X$^0$), negatively charged (X$^-_T$) excitons, as well as low energy states (L) is indicated with different colors after deconvolution using Lorentzian functions.

## II. RESULTS AND DISCUSSION

Atomically thin WSe$_2$ flakes are prepared by mechanical exfoliation of a bulk crystal (2D semiconductors) on polydimethylsiloxane (PDMS). WSe$_2$ monolayers are identified by optical contrast microscopy and subsequently transferred on silicon (Si) wafers covered by silicon dioxide (SiO$_2$/Si, 290 nm-thick). PL and Raman spectroscopy are utilized to verify the number of layers after transfer (see Figs. S1, S2 in the Supplemental Material (SM) [20]). A characteristic optical image of 1L WSe$_2$ on SiO$_2$/Si is shown in Fig. 1(a) and its corresponding Raman spectrum is presented in Fig. 1(b). In the limit of 1L, the two degenerate first order Raman modes E′ and A′$_1$ located approximately at 250 cm$^{-1}$ have about 11 cm$^{-1}$ difference from the second order 2LA(M) mode [21–23]. Fig. 1(c) shows a typical PL spectrum of untreated 1L-WSe$_2$ collected at 78K, where X$^0$, X$^-_T$ and low energy (L) excitons (stemming from an ensemble of dark, localized and phonon replica states) are evident. The 34 meV energy difference between X$^0$ and X$^-_T$ supports the negative total charge of the trion peak [24] and verifies that our samples are naturally electron-doped. We fit the data with three Lorentzian functions resulting in large values of coefficient of determination in the total fit, R$^2$>0.99.

Following optical characterization of the untreated samples, monolayers are first pumped down to ~10 Torr and subsequently exposed to 120 Torr gaseous chlorine (Cl$_2$) environment. A 248 nm UV excimer laser, with a pulse duration of 20 ns and a repetition rate of 1 Hz (1 pulse per second) is used to controllably irradiate the samples in the presence of Cl$_2$. The samples are removed from the chamber after each irradiation pulse for low temperature optical characterization.

To evaluate the impact of the photochlorination process on the carrier density of 1L-WSe$_2$ we use low-temperature PL spectroscopy (see Fig. 2(a)). A laser with a wavelength of 543 nm (2.28 eV) is used to excite above the optical gap of 1L-WSe$_2$ while maintaining a fixed power of 60 μW to avoid thermal effects. Interestingly, the energy position of X$^0$ redshifts after 1 second exposure and thereafter it saturates (see Fig. 2(b)). The opposite trend is noticed for charged excitons showing a slight blueshift. It is well established by several studies using electrostatically-gated samples that in 1L-WSe$_2$ X$^-_T$ and X$^+_T$ have considerably different binding energies, owing to the different effective mass between electrons and holes [24,25]. In our experiments, the energy difference between X$^0$ and X$_T$ is rapidly reduced from 34 meV to 20 meV after 1 second of photochlorination time (see Fig. 3(c)), in perfect agreement with reported binding energies of X$^-_T$ and X$^+_T$ [24,25]. Note that the same behavior is observed in samples where 1L-WSe$_2$ is transferred on ~150 nm thick hexagonal boron nitride (h-BN) showing narrower emission linewidths (see Fig. S3 in the SM [20]) [26–28]. It is important to mention that neutral biexcitons also lie very close to X$^0$ (15 meV binding energy) [24,29,30]. However, the assignment to neutral biexcitons is excluded since excitation power experiments indicate a linear dependence of the emission (see Fig. S4 in the SM [20]) and not quadratic. We further define ρ as the intensity ratio of X$^0$ over either negatively (X$^-_T$) or positively (X$^+_T$) charged excitons. We observe that ρ monotonically decreases as a function of photochlorination time, indicating that emission from X$^+_T$ is the dominant recombination channel after photochlorination (see Fig. S5 in the SM [20]). The above observations suggest a rapid and complete depletion of excess electrons in the pristine sample and additional hole doping in the crystal even only after 1 second of photochlorination time.

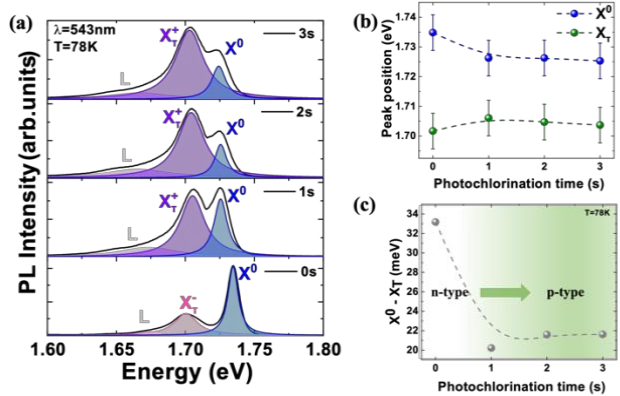

FIG.2. (a) Evolution of the 1L WSe$_2$ PL spectra at 78K, with increasing photochlorination time. (b) Peak position of $X^0$ and $X_T$ with increasing photochlorination time. (c) Energy difference between $X^0$ and $X_T$ with increasing photochlorination time.

We now comment on the experimental validation of the presence of chlorine species in the monolayer lattice after photochlorination. The XPS core level spectra may provide substantial input regarding the chemical state of the studied 1L-WSe$_2$ samples. The binding energy of the Se-$3d$ states of the studied samples WSe$_2$ agrees with previous reports [31]. The W-$4f$ peak overlaps with the strong O-$2s$ of the substrate. The most interesting information from the core levels comes from the comparison of the Cl-$2p$ from the photochlorinated 1L-WSe$_2$ of the present study with the photochlorinated WS$_2$ from previous reports [18]. The Cl-$2p$ peaks of the photochlorinated WS$_2$ are consistent with adsorbed Cl$_2$ [32], or to volatile transition metal halide molecules such as TiCl$_4$ [33] and ZrCl$_4$ [34]. The latter indicates that Cl might be chemically bonded with the transition metal (W) atoms at defects sites. The relevant Cl-$2p$ peaks are also present in the photochlorinated WSe$_2$ (Fig. 3) and show similarities to those of solid metal halides such as MoCl$_2$ [35], ZnCl$_2$ [36], and CdCl$_2$ [36]. In particular, ZnCl$_2$ and CdCl$_2$ may form 2D structures [37], which are structurally compatible to WSe$_2$. Thus, Cl can dope WSe$_2$ by being incorporated into its lattice in Se sites or in Se-vacancy sites as it is revealed by the following DFT calculations.

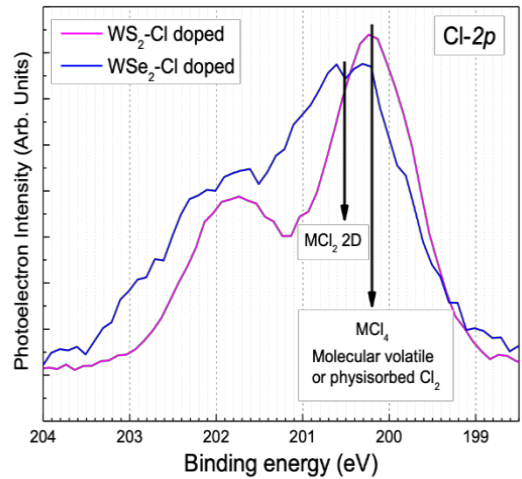

FIG.3. Cl-$2p$ core levels for photochlorinated WSe$_2$ and WS$_2$.

We perform DFT calculations with atomic and molecular chlorine adsorbed on monolayers, both with and without vacancies, to gain deeper insight into the electronic properties of photochlorinated 1L-WSe$_2$. Simulation cells of different sizes are used to examine the effects of different vacancy and chlorine concentrations, which turn out to be non-essential: when chlorine concentration increases, phenomena described in the following become more pronounced but remain qualitatively the same. Fig. 4 illustrates the main theoretical findings which agree with the experimental results. In Fig. 4(a), the well-known band structure of the pristine monolayer is shown. In Fig. 4(b), the effective band structure (EBS), after unfolding [38,39], for 1L-WSe$_2$ with Se-vacancy (V$_{Se}$) defect is presented. It is evident that the presence of V$_{Se}$ turns an intrinsic semiconductor into an electron-doped one, with defect states appearing close to the conduction band minimum (CBM) when comparing the band structures of perfect crystalline and defected monolayers. Formation energies for $V_{Se}$ were calculated close to 1.8 eV for several concentrations (see Table S1 in the SM [20]). In Fig. 4(c), the EBS for atomic chlorine adsorbed on Se-vacancy is presented, showing a certain degree of passivation of the $V_{Se}$ defect states close to the CBM. The EBS for molecular chlorine adsorbed on $V_{Se}$ in Fig. 4(d) clearly shows p-type doping with defect states appearing close to the valence band maximum (VBM). Adsorption energies on the vacancy site are around -1.2 eV for atomic Cl and -2.0 eV for $Cl_2$, making molecular chlorine adsorbed on Se-vacancy the most energetically favorable configuration (exact values of energies for several concentrations can be found in the SM [20], Table S2). The additional adsorption energy of -0.8 eV in the case of $Cl_2$-filled $V_{Se}$ suggests strong bonding between one Cl atom and an existing Cl-filled $V_{Se}$. For all simulation cells, EBSs (see Fig. S6 in the SM [20]) show that $Cl_2$ adsorbates on Se-vacancies result in p-type doping, in good agreement with experimental data.

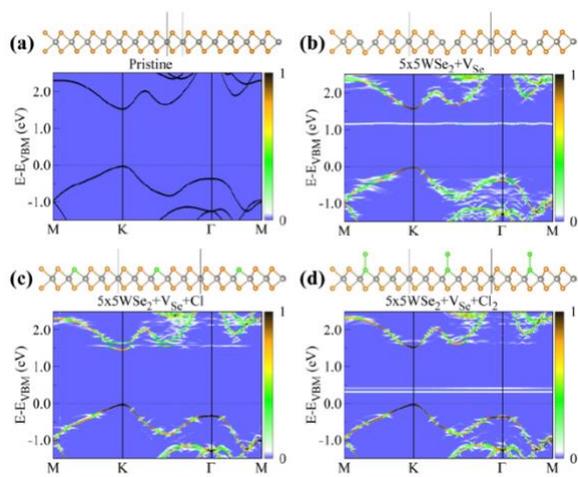

FIG.4. a) Side view of the pristine 1L-WSe$_2$ unit cell and the corresponding band structure. (b) 1L-WSe$_2$ 5x5 simulation cell with a Se-atom vacancy, V$_{Se}$, and its effective band structure. (c) and (d): Similar to (b) but with atomic Cl and Cl$_2$, respectively, adsorbed on V$_{Se}$.

In addition to being consistent with experiments, our DFT calculations consider other scenarios. We examine chlorine adsorption on vacancy-free monolayers. We find that atomic chlorine adsorption over a perfect $WSe_2$ surface is energetically unfavorable (positive adsorption energy, more than 0.5 eV, corresponding EBS Fig. S7(a)) while molecular chlorine binds very weakly to the monolayer, with energies between -0.14 and -0.03 eV (see Table S3 in the SM [20], corresponding EBS Fig. S7(b)). Binding of $Cl_2$ becomes even weaker if zero-point vibrational motion and entropy effects are considered, resulting in adsorption energy increase of the order of a tenth of one eV. In addition, physisorption of $Cl_2$ on perfect monolayers results in n-doped samples with electronic states close to the CBM (see Fig. S7(b) in the SM [20]) which is not in agreement with experiments. However, we emphasize that a perfectly structured, vacancy-free monolayer is not a realistic scenario [40] and the presence of $V_{Se}$ is necessary to favor adsorption of Cl-species (either atomic or molecular). Thus, DFT calculations indicate that 1L-$WSe_2$ samples, which are intrinsically n-doped due to a finite density of $V_{Se}$, exhibit optical properties consistent with p-type doping after photochlorination.

In the last part, we examine the reversibility of the photochlorination process. We use a 473 nm (≈2.62 eV photon energy) continuous wave (CW) laser with a power density of 380 mW/cm² and we scan the photochlorinated samples for 180 sec at room temperature, in ambient conditions. Subsequently, we place the samples back in the cryostat for examination of the PL emission. Fig. 5(a) compares PL spectra between pristine 1L-$WSe_2$, after 10 sec photochlorination and after CW treatment. While $X^+_T$ dominates the PL spectra after photochlorination, it is evident that $X^0$ only partially recovers after CW treatment with $X^+_T$ still present. Based on the emission intensities shown in Fig. 5(b) we demonstrate that only ~15% of the process is reversed. The result comes as no surprise if we consider the DFT calculations that predict a gain in the binding energy of $Cl_2$ bonding of the same order of magnitude. Consequently, in a rough estimation the photon energy of 2.62 eV of the CW laser is barely sufficient to remove chlorine species from the photochlorinated samples. One possible scenario of the observed weak reversibility is a partial transition from molecular adsorption (p-type doping, Fig. 4(d)) to an atomic adsorption (passivated case, Fig. 4(c)).

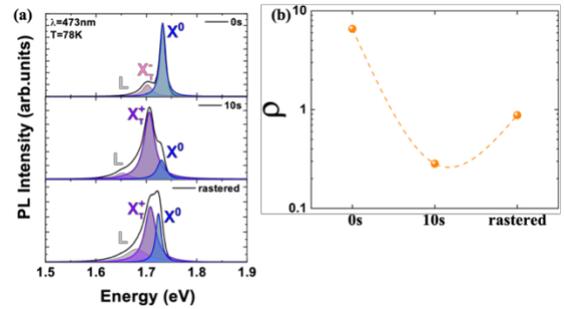

FIG.5. (a) PL spectra of pristine, photochlorinated and rastered 1L-$WSe_2$. (b) Intensity ratio, $\rho$, of the 3 different cases.

### III. CONCLUSIONS

Based on strong evidence, we propose that the semiconductor properties of 1L-$WSe_2$ undergo a switch from n-type to p-type after photochlorination in a $Cl_2$ precursor. The transition is verified by low temperature PL spectroscopy where the trion emission switches from negatively to positively charged species with increasing photochlorination steps. XPS verifies the presence of chlorine species in the monolayers and DFT calculations predict p-type doping when chlorine is adsorbed on Se-site vacancies. Due to the relatively strong adsorption of both molecular and atomic chlorine in the lattice the process is only weakly reversible. Our results suggest that photochemical techniques have the potential to tune the optical response of 2D materials, making them attractive for optoelectronics and photonics.

## IV. MATERIALS AND METHODS

### A. Sample preparation

Monolayers WSe$_2$ were prepared by mechanical exfoliation from a bulk crystal (2D semiconductors). During the exfoliation an adhesive tape or "Scotch-tape" was used. A repetitive folding-unfolding process was carried out in order to spread the bulk crystal along the tape in thinner flakes. After that, the sticky side of the scotch tape was placed on the substrate to force material to move from the upper parts of WSe$_2$ flakes to the SiO$_2$/Si substrate. Steadily and carefully, we pealed the substrate off the scotch-tape leaving the monolayer flake on it.

### B. Photochlorination process

The pristine WSe$_2$ monolayers were irradiated by a KrF excimer UV laser of 248 nm. The pulse duration was 20ns and the repetition rate was 1Hz. A beam homogenizer was used to offer uniform exposure to irradiation. The monolayers were placed into a vacuum chamber in Cl$_2$ environment of 120 Torr gas pressure. A parametric study of laser power (P) and number of pulses ($N_p$) was executed in order to find the ideal conditions in the case of 1L-WSe$_2$. The ideal laser power was P=10 mJ/cm$^2$ and all the experiments with different number of pulses, that is photoreaction time, were executed with this constant laser power.

### C. Optical spectroscopy

Optical spectroscopy was performed by a micro-Reflectivity and a micro-PL setup. The laser beam was focused by a Mitutoyo 50x (NA:0.42) lens onto the sample. The laser beam size was about ~1 μm. The sample was located into continuous-flow cryostat (ST500, Jannis). We controlled the position of the sample through a *XYZ* mechanical translation stage (PT3, Thorlabs). In parallel, the whole process of excitation was controlled through a CCD (charge-coupled device) optical setup. The optical setup consists of a short pass filter to eliminate the noise at higher wavelengths and a long pass filter, through which passes the emitted PL signal, to diminish the emission of the laser.

### D. X-ray photoelectron spectroscopy

Core-level and valence band XPS were acquired in a KRATOS Axis Ultra DLD system equipped with a monochromated AlKa x-ray source, a hemispherical sector electron analyzer and a multichannel electron detector. The reported spectra were acquired without any surface cleaning, because the monolayer nature of the samples did not endure any mild sputter etching [41]. The XPS measurements were acquired using 20 eV pass energy resulting in a full width at half maximum of the Ag-3d$^{5/2}$ peak of less than 500 meV. Spectral shifts due to charging of the surface were evaluated and subtracted based on the spectral positions of the C-1s peak of adventitious Carbon [42].

### E. Theoretical calculations

DFT calculations were performed with VASP [43]. The projector augmented-wave (PAW) method for core electrons and nuclei [44,45] and the generalized gradient approximation of Perdew−Burke−Ernzerhof (PBE) for the exchange-correlation functional [46] were employed. Wave functions were expanded on a plane-wave basis set with a cutoff of 450 eV. The k-point sampling was set at 37x37x1 for the primitive unit cell of WSe$_2$ and appropriately scaled for supercells. We let the lattice vectors and ions to fully relax with a convergence criterion of 0.01 eV/Å for ions and 10$^{-5}$ eV for electrons, which became 10$^{-8}$ eV for the band structures. Results for 1L WSe$_2$ are in agreement with previous works (e.g. [47]). In order to examine vacancies and chlorine adsorption, 3×3, 4×4, and 5×5 simulation cells (SCs) were considered, corresponding to approximately 11%, 6%, and 4%, defect concentrations, respectively. Their effective band structures (EBS) were calculated from projecting the eigenstates of the SCs to the primitive unit cell (PC) of the pristine case [48], in a way that allows for a one-to-one comparison of the band structures of pristine and doped cases. Both band structure and effective band structures are presented with color as spectral functions as described in [38], where our method for unfolding the band structure from the SC Brillouin zone (BZ) to the PC BZ is outlined. Our band unfolding method for van der Waals heterostructures, applied here for WSe$_2$ monolayers on h-BN to show that results are similar to free-standing sample (Fig. S8 in the SM [20]), is described in [39].

### ACKNOWLEDGEMENTS

E.K acknowledge support by the European Regional Development Fund of the European Union and Greek national funds through the Operational Program Competitiveness, Entrepreneurship and Innovation, under the call RESEARCH – CREATE – INNOVATE (SMARTPACK), Innovation (H.F.R.I.) under the '4th Call for H.F.R.I. Scholarships to PhD Candidates' project No: 9231. I.D. acknowledge support by the European Union's Horizon 2020 research and innovation program through the project NEP, EU Infrastructure, GA 101007417 – INFRAIA-03-2020 and by the synergy grant SPIVAST funded by FORTH. Funding by the Hellenic Foundation for Research and Innovation (H.F.R.I.) under the 'First Call for H.F.R.I. Research Projects to support Faculty members and Researchers and the procurement of high-cost research equipment grant' project No: HFRI-FM17-3034 (G. Kio.) and project No: HFRI-FM17-1303 (G. V. and G. Kop.) is acknowledged.
First-principles calculations were performed with computational time granted from the Greek Research and Technology Network S.A. (GRNET S.A.) in the National HPC facility ARIS under Projects NANOCOMPDESIGN and NANOPTOCAT. K.W. and T.T. acknowledge support from the JSPS KAKENHI (Grant Numbers 20H00354, 21H05233 and 23H02052) and World Premier International Research Center Initiative (WPI), MEXT, Japan.

# Supplemental material

# Electron density control in WSe$_2$ monolayers via photochlorination

E. Katsipoulaki[1,2], G. Vailakis[1,3], I. Demeridou[1], D. Karfaridis[4], P. Patsalas[4], K. Watanabe[5], T. Taniguchi[6], I. Paradisanos[1], G. Kopidakis[1,3], G. Kioseoglou[1,3] and E. Stratakis[1,2]

[1]Institute of Electronic Structure and Laser, Foundation for Research and Technology-Hellas, Heraklion, 71110, Crete, Greece
[2]Department of Physics, University of Crete, Heraklion, 71003, Crete, Greece
[3]Department of Materials Science and Technology, University of Crete, Heraklion, 71003 Crete, Greece
[4]Department of Physics, Aristotle University of Thessaloniki, Thessaloniki, 54124, Greece
[5] Research Center for Electronic and Optical Materials, National Institute for Materials Science, 1-1 Namiki, Tsukuba 305-0044, Japan
[6] Research Center for Materials Nanoarchitectonics, National Institute for Materials Science, 1-1 Namiki, Tsukuba 305-0044, Japan

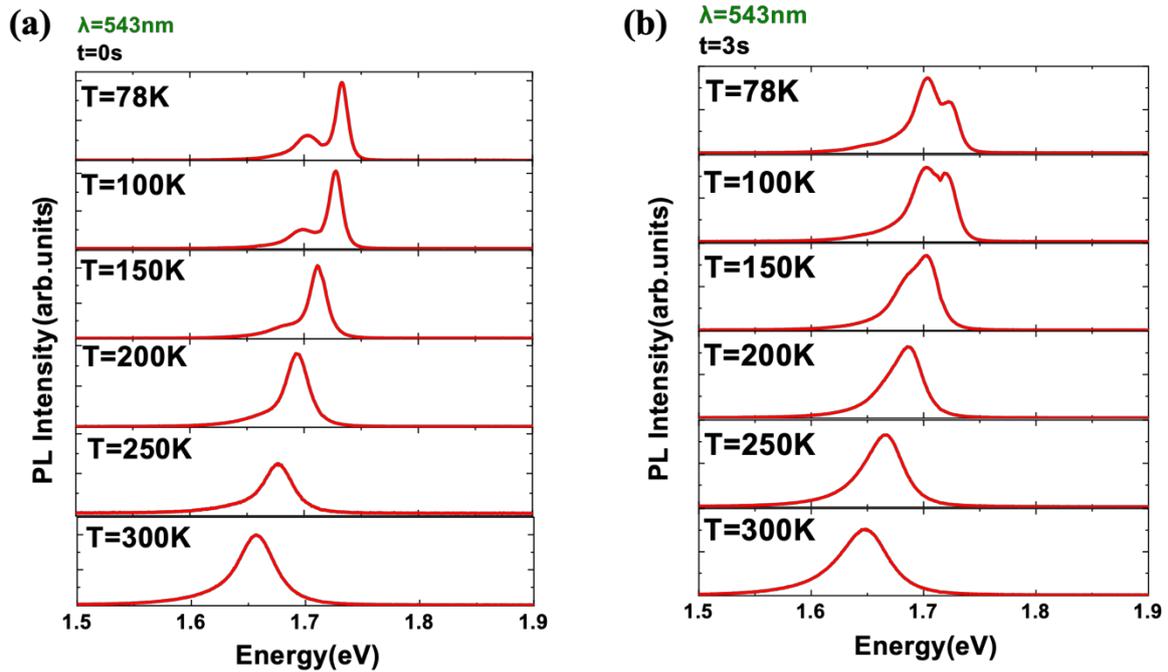

FIG. S1. (a) Temperature-dependent PL spectra for the pristine sample from 78K to 300K. (b) Temperature-dependent PL spectra for the photochlorinated sample (t=3s) from 78K to 300K.

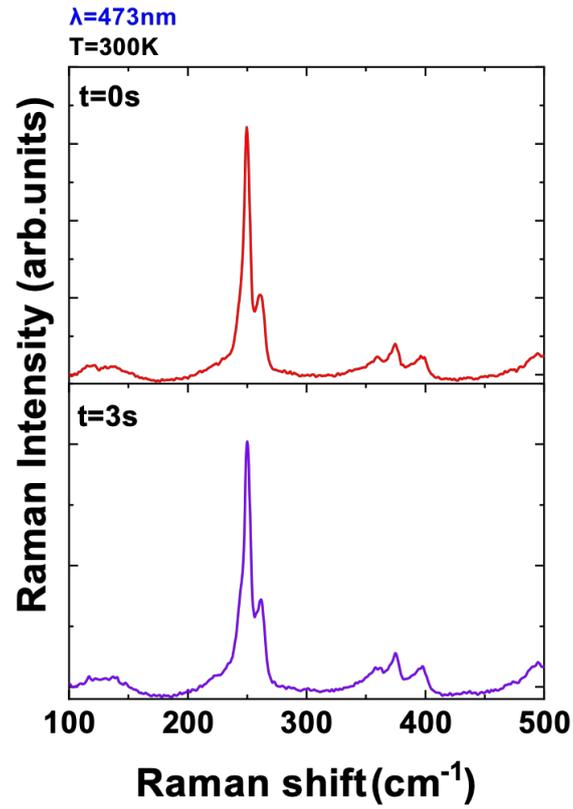

FIG. S2. Room temperature Raman spectra collected with 473nm laser excitation for the pristine (red line) and the photochlorinated (magenta line) sample.

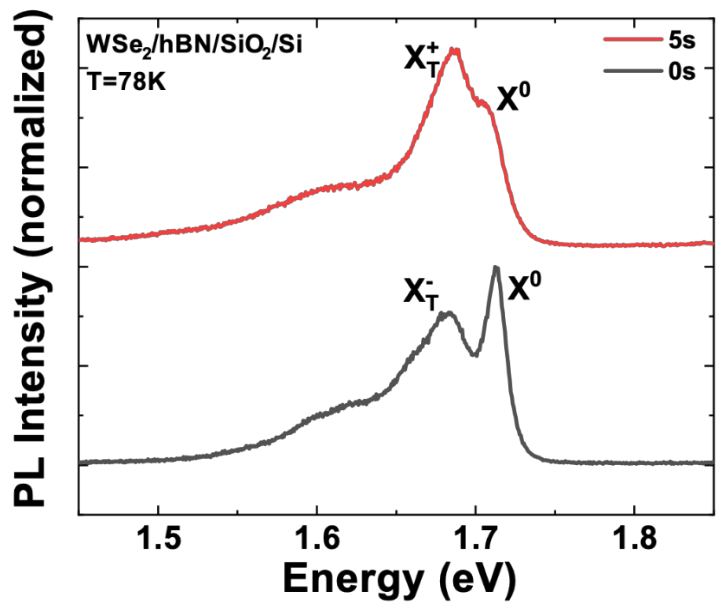

FIG. S3. PL spectra of 1L-WSe$_2$/hBN at 78 K collected with 635 nm laser excitation for the pristine (grey line) and the photochlorinated (red line) sample.

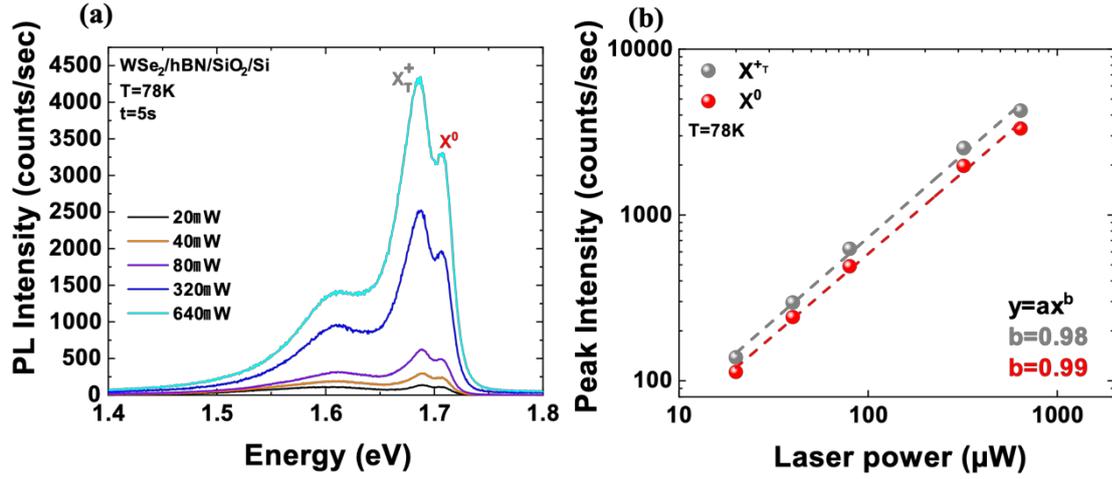

FIG. S4. (a) PL spectra for different laser powers at 78 K with 543 nm excitation wavelength for the photochlorinated sample (t=3s). (b) Double-logarithmic plot of $X_T^+$ integrated intensity as a function of laser power.

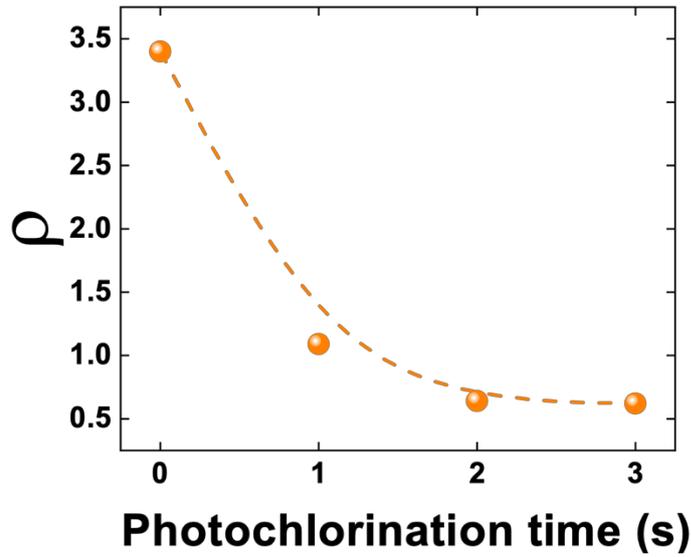

FIG. S5. Intensity ratio ($\rho$) of the $X^0$ over $X_T$ intensity as a function of the photochlorination time.

| Vacancy Formation Energy | | |
|---|---|---|
| Conc. | Supercell | $WSe_2$ |
| 4% | 5x5 | 1.783 eV |
| 6% | 4x4 | 1.786 eV |
| 11% | 3x3 | 1.793 eV |

TABLE S1. The formation energy ($E_f = E_{(WSe_2+V_{Se})} + \mu_{(Se)} - N_{(WSe_2)}E_{(WSe_2)}$) of vacancies in different supercells (or concentrations).

| | Adsorption Energies on Chalcogen Vacancies | |
|---|---|---|
| Conc. | $WSe_2$ $V_{Se}$ | |
| | Cl | $Cl_2$ |
| 4% | -1.190 eV | -2.029 eV |
| 6% | -1.226 eV | -2.011 eV |
| 11% | -1.245 eV | -1.956 eV |

TABLE S2. The adsorption energy ($E_{ads} = E_{(WSe_2+V_{Se}+Cl_n)} - E_{(WSe_2+V_{Se})} - \mu_{(Cl_n)}$) of Cl atom and molecule on Se vacancies for 1L-$WSe_2$.

| | | Adsorption Energies | |
|---|---|---|---|
| Conc. | Position | $WSe_2$ | |
| | | Cl | $Cl_2$ |
| 4% | on top Se | 0.554 eV | -0.140 eV |
| | on top W | 0.648 eV | -0.026 eV |
| | on top h | 0.517 eV | -0.038 eV |
| 6% | on top Se | 0.547 eV | -0.143 eV |
| | on top W | 0.657 eV | -0.030 eV |
| | on top h | 0.518 eV | -0.044 eV |
| 11% | on top Se | 0.570 eV | -0.136 eV |
| | on top W | 0.697 eV | -0.030 eV |
| | on top h | 0.540 eV | -0.043 eV |

TABLE S3. The adsorption energy ($E_{ads} = E_{(WSe_2+Cl_n)} - E_{(WSe_2)} - \mu_{(Cl_n)}$) of Cl atom and molecule on top of selenium (Se), tungsten (W) and middle of the hexagon (h) of 1L-$WSe_2$.

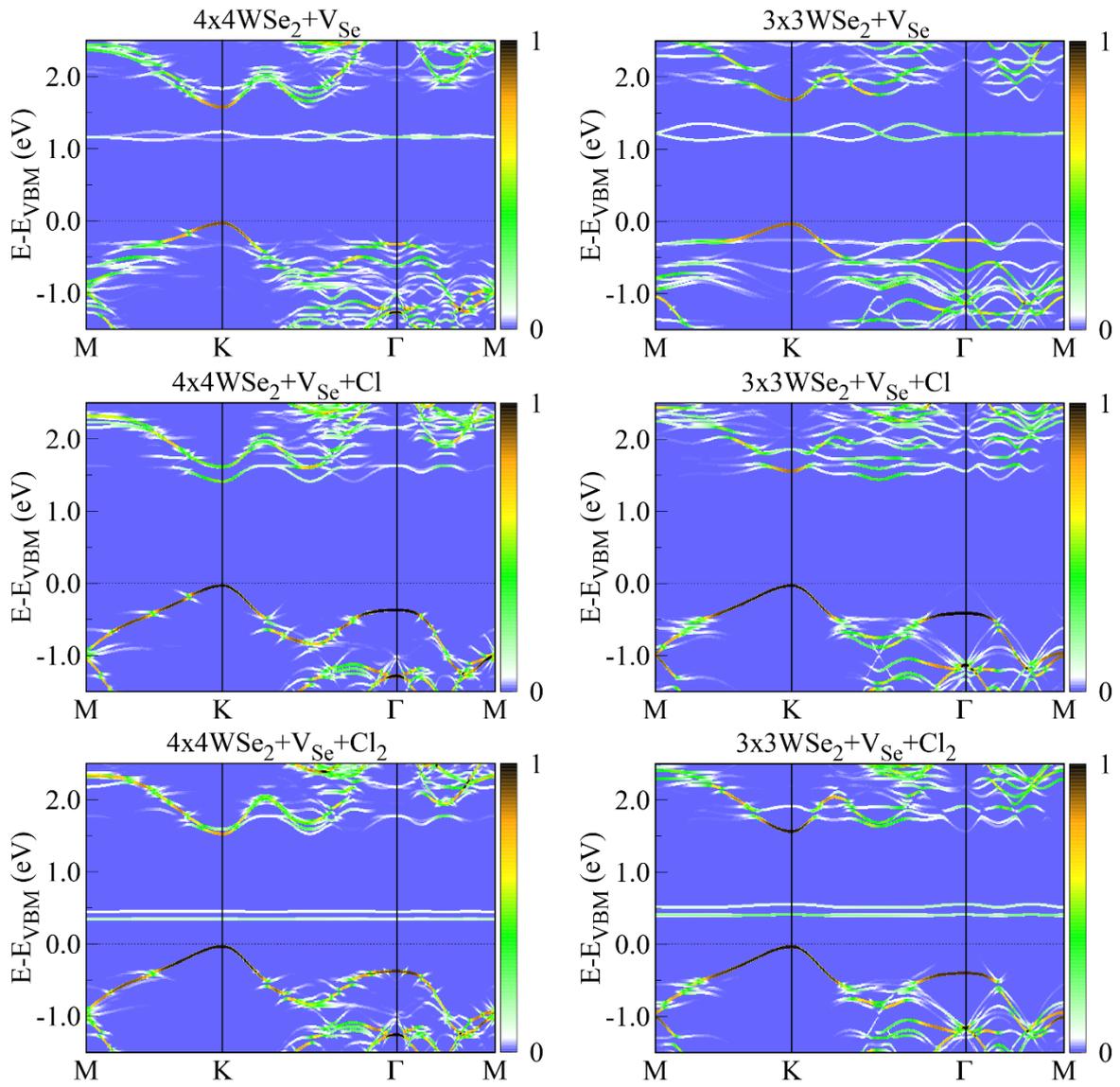

FIG. S6. Effective band structures for 4x4 and 3x3 WSe2 supercells with Se-vacancy.

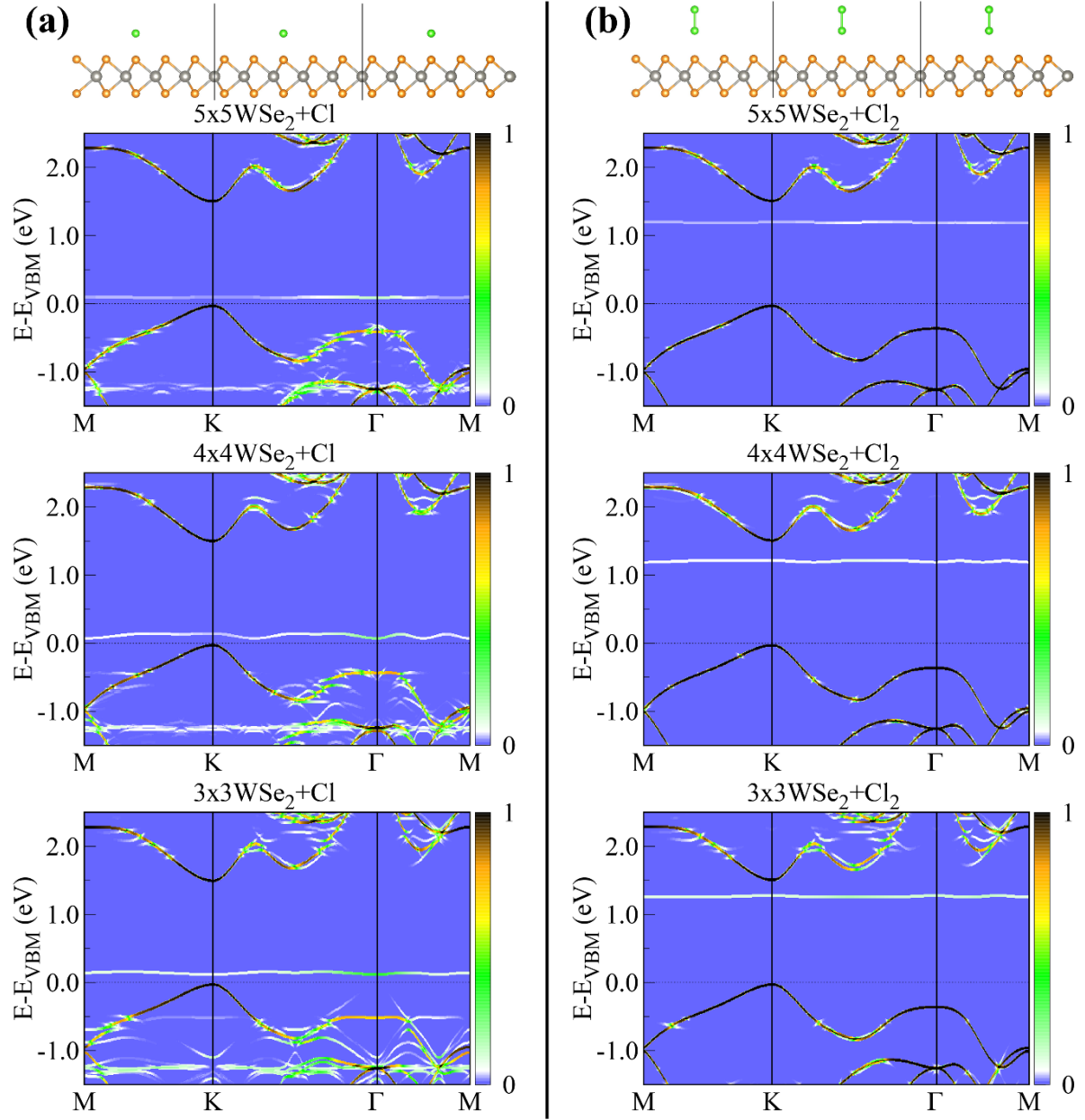

FIG. S7. Effective band structures for adsorbed Cl (a) and Cl$_2$ (b) on vacancy free WSe$_2$.

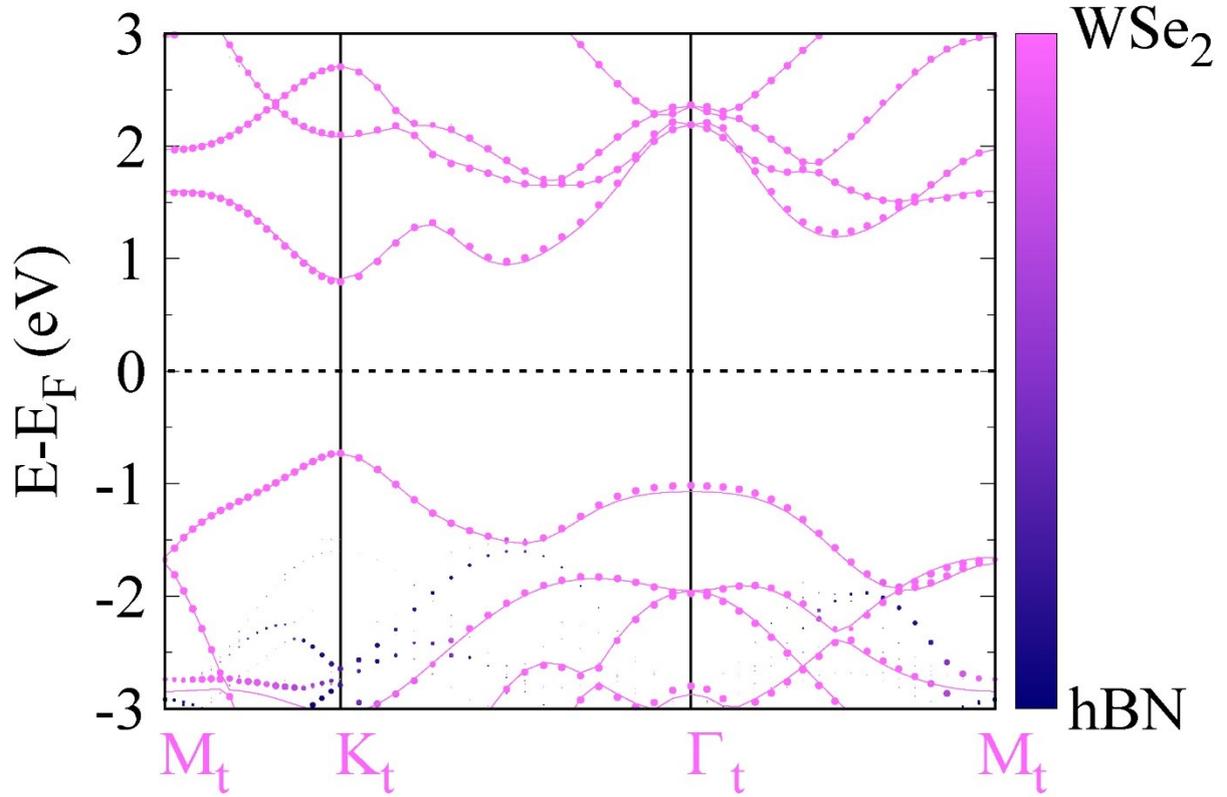

FIG. S8. Effective band structure of WSe$_2$ on-top of hBN with a twist-angle of 19.10°. The path follows the MKΓM of the 1$^{st}$ Brillouin Zone for the primitive cell of WSe$_2$. The spectral weight of each electronic state is proportional to the radius and with color we denote the position of each state in the direction perpendicular to the monolayers from WSe$_2$ (magenta) to hBN (blue). The band structure of the free-standing WSe$_2$ is illustrated with the solid magenta line. By construction, strain due to lattice mismatch in the simulation cell is minimal (0.14% expansion for WSe$_2$ and 0.14% compression for hBN).